\documentclass[acmlarge]{acmart}
\usepackage{subfigure}
\usepackage{tabularx}

\setcopyright{acmcopyright}
\copyrightyear{2023}
\acmYear{2023}

\acmConference[ISMSI]{2023 7th International Conference on Intelligent Systems, Metaheuristics & Swarm Intelligence (ISMSI 2023)}{April 23–24, 2023,}{Virtual Event, Malaysia}
\acmDOI{10.1145/3596947.3596966}
\acmBooktitle{2023 7th International Conference on Intelligent Systems, Metaheuristics \& Swarm Intelligence (ISMSI 2023), April 23–24, 2023, Virtual Event, Malaysia}

\begin{document}

\title{Machine Learning for Socially Responsible Portfolio Optimisation}

\author{Taeisha Nundlall}
\email{taeishanundlall@gmail.com}
\affiliation{%
  \institution{University of Johannesburg}
  \city{Johannesburg}
  \country{South Africa}}

\author{Terence L. van Zyl}
\email{tvanzyl@gmail.com}
\affiliation{%
  \institution{University of Johannesburg}
  \city{Johannesburg}
  \country{South Africa}}

\renewcommand{\shortauthors}{Nundlall and van Zyl}

\begin{abstract}
Socially responsible investors build investment portfolios intending to incite social and environmental advancement alongside a financial return. Although Mean-Variance (MV) models successfully generate the highest possible return based on an investor’s risk tolerance, MV models do not make provisions for additional constraints relevant to socially responsible (SR) investors. In response to this problem, the MV model must consider Environmental, Social, and Governance (ESG) scores in optimisation. This study implements portfolio optimisation for socially responsible investors based on the prominent MV model. The amended MV model allows SR investors to enter markets with competitive SR portfolios despite facing a trade-off between their investment Sharpe Ratio and the average ESG score of the portfolio.
\end{abstract}

%




\maketitle

\section{Introduction}
\label{section:intro}
Given the climate crisis and the ongoing socioeconomic issues, more investors are looking to participate in the markets through socially responsible investment (SRI) \cite{chen2021social}  decisions. Socially responsible (SR) investors build investment portfolios to initiate social and environmental advancement alongside a financial return \cite{branch2019guide}. The availability of Environmental, Social and Governance (ESG) ratings\footnote{Also known as ESG issues, are non-financial components that are investigated to rank and compare the sustainability and ethical impact of various securities} provides valuable information that can be incorporated into SRI decisions, where both moral value and economic value are of concern. During SRI asset allocation, an additional trade-off \cite{de2021esg} exists between portfolio returns, risk and ESG-ratings (i.e. Sharpe Ratio and ESG). Although Mean-Variance (MV) \cite{Markowitz1952} models successfully generate the highest possible return based on the investor's risk tolerance, MV models do not make provisions for additional constraints relevant to SR investors~\cite{paskaramoorthy2020framework,skeepers2021ma,freeborough2022investigating}. In response to this problem, we propose to extend the MV Model to include ESG-rating optimisation. A tri-criterion (i.e. mean, variance, ESG-rating) portfolio optimisation model allows investors to align investment decisions with moral values better, benefiting from such decisions' economic value.

\citet{Vo2019} provides a Deep Responsible Investment Portfolio (DRIP), a portfolio optimization model that uses deep learning and incorporates ESG-ratings. Stock prices and ESG-ratings are inputs in a Multivariate Bi-directional Long Short-Term Memory (LSTM) neural network to predict stock returns. Deep reinforcement learning is used to re-train neural networks and re-balance the portfolio. The DRIP framework achieved competitive financial performance and better social impact than traditional portfolio models, sustainable indexes and funds. This literature is relevant to our research and provides useful guidance on incorporating ESG-ratings into deep learning models. 

This study uses machine learning~\cite{muthivhi2022fusion,laher2021deep,van2021parden} to calculate optimal SRI portfolios to evaluate the effect \cite{ouchen2022esg} of considering ESG-ratings in portfolio optimisation. Exchange-traded funds (ETF) closing prices are forecasted using a random forest (RF) regression algorithm to produce out-of-sample data for use during asset allocation. This study uses an amended MV model \cite{Vo2019} to predict portfolio weightings to optimise the portfolio Sharpe Ratio \cite{Sharpe1966} and ESG-ratings. The research value for this study is a look at the effect of ESG-ratings as a third criterion in portfolio optimisation when using machine learning to forecast ETF returns.

This paper is structured as follows. In Section \ref{section:intro}, a brief introduction to the study is presented. In Section \ref{section:method}, the research methodology is discussed in detail. In Section \ref{section:results}, results from the study's machine learning and portfolio optimisation components are discussed, respectively. Section \ref{section:conc} presents the conclusions of this study.

\section{Method} \label{section:method}
\begin{figure}[htb]
    \centering
\includegraphics[width=\columnwidth]{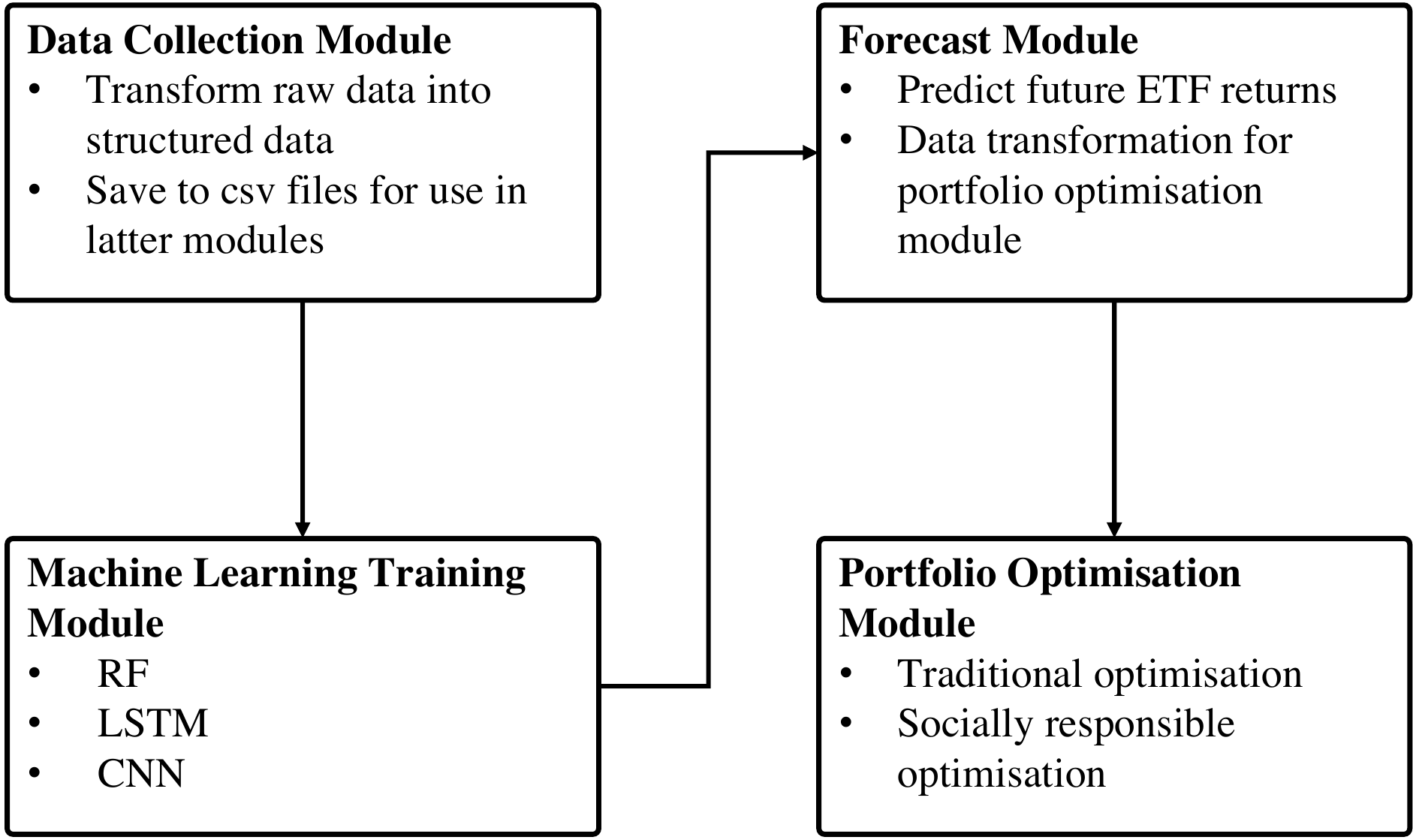}
    \caption{A diagram expressing the main idea of the study}
    \label{fig:studydiag}
\end{figure}

This study consists of two main components: the use of machine learning to forecast the closing prices of the ETF and the optimisation of a portfolio using traditional mean-variance and the ESG-MV methods proposed in \citet{Vo2019}. Details on the research methodology are discussed in this section, and the general structure of the methodology is depicted in Figure \ref{fig:studydiag}. 
\subsection{Datasets} \label{subsection:dataset}
\begin{figure}[htb]
    \centering
\frame{\includegraphics[scale=0.25,width=\columnwidth]{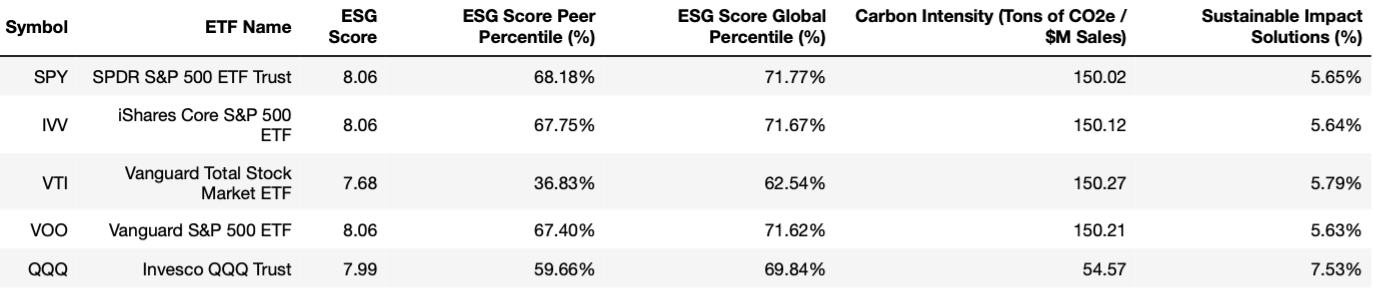}}
    \caption{ESG dataset released by VettaFi}
    \label{fig:ESGSCORESDB}
\end{figure}

VettaFi owns databases containing various financial datasets related to Exchange-Traded Funds (ETF). ETFs are a relatively new pooled investment security, with SPDR S\&P 500 ETF Trust being the first ETF available on US exchanges, debuting as recently as 1993. VettaFi has gone one step further than their industry competitors to include ESG data for all ETFs captured on their system and has been made available for download to the public. We have downloaded the ESG data of U.S. ETFs traded on U.S. exchanges. Other ETF data are included in the VettaFi database, such as returns, leverage status, and year-to-date ETF prices. The dataset structure used can be seen in Figure \ref{fig:ESGSCORESDB}. This study uses the ESG scores from this dataset, ranging from 0 to 10. The other features of the dataset are redundant, as they are merely sub-components of the ESG score.

\begin{figure}[htb]
    \centering
\frame{\includegraphics[width=\columnwidth]{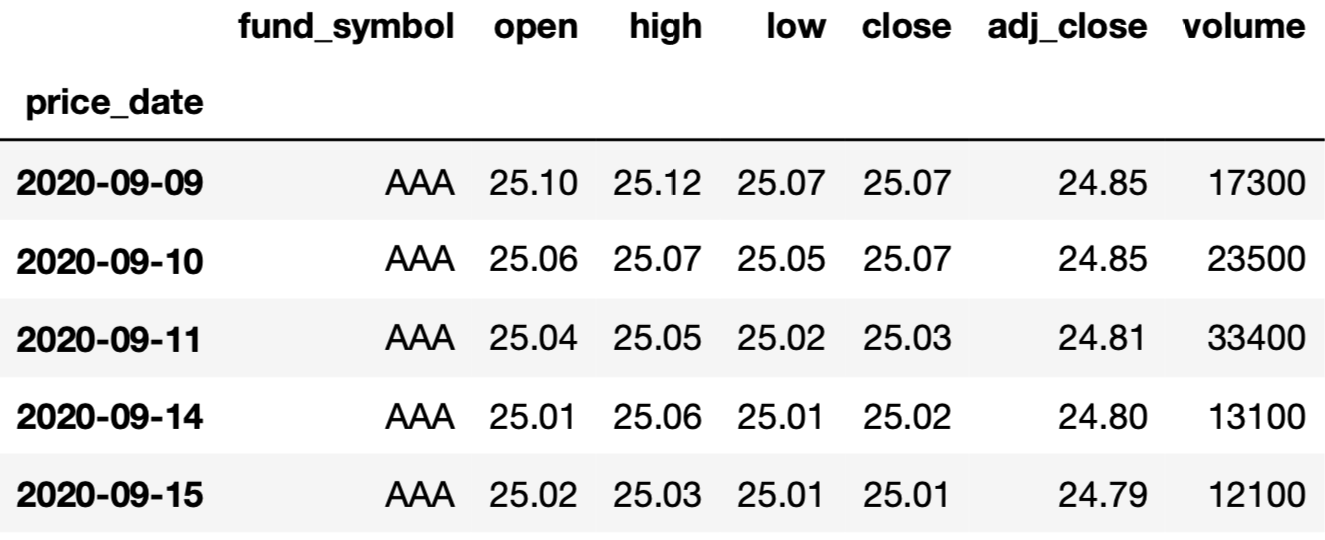}}
    \caption{ETF financial data released by Yahoo Finance}
    \label{fig:ETFPrices}
\end{figure}

Yahoo Finance has released datasets containing U.S ETFs and their historical prices from 1993-01-29 to 2021-11-30. The version used in this study is related to the financial values of November 2021. The structure of the dataset used can be seen in Figure \ref{fig:ETFPrices}.
More than 80\% of the historical data belong to trading days between 2011-11-30 and 2021-11-30. This is due to the increased number of ETFs available on the US exchange during this period. Our study limits the use of historical data to trading days between 2011-11-30 and 2021-11-30 to avoid excessive back-filling for the 80\% of the ETFs that were non-existent before 2011-11-30.
Before supplying the Yahoo Finance dataset to the portfolio optimisation component of the study, data related to ETFs not contained in the VettaFi dataset is removed to only consider ETFs with recorded ESG ratings. This reduces the number of ETFs in the dataset from 1644 to 1358.

\subsection{Study Procedure}
After performing the data preprocessing as detailed, the dataset is filtered to consider the adjusted closing values for one ETF at a time. The dataset is split 4:1 into training and testing datasets. 

The historical data of one ETF is used to train and test each of the three machine learning models; random forest, LSTM, and Convolutional Neural Network (CNN). Using other ETFs, re-runs of the prediction models are performed to determine if the performance of the models is consistent. ETFs considered in the experiment are Betashares Australian High-Interest Cash ETF (AAA), SPDR S\&P 500 ETF Trust (SPY), and Innovator US Equity Ultra Buffer ETF - October (UOCT).

Deep learning models have been constructed using the Adam optimiser, ReLu activation functions, and the mean absolute error (MAE) loss function. The CNN also made use of max pooling. The RF model is constructed with 10000 n\_esimators.

All predictive models' mean absolute scaled error (MAE) MASE and root mean squared scaled error (RMSSE) are computed and evaluated to determine the best-performing model during training. The random forest model outperforms the other models and is further discussed in Section \ref{section:results}. The RF model forecasts daily closing prices for the two months following 2021-11-30. The historical and forecasted adjusted closing values, along with the ESG scores from the VettaFi dataset, were stored in a new dataset for the utility of the portfolio optimisation module. 

For portfolio optimisation \cite{Pai2018}, we used historical data from 2011-12-01 to 2021-11-30, 10 years, and included our forecast data from the forecasting component of the study. As done in \cite{Vo2019}, portfolio optimisation, for both MV and ESG-MV portfolio optimization, is done using the Sequential Least SQuares Programming (SLSQP) optimisation method.

No bounds have been placed on ESG scores in data pre-processing and portfolio optimisation modules. This is so that the entire market of 100 shares is considered in both the MV and the ESG-MV models. This will allow us to study the optimal weights allocated to assets with high and low ESG scores.

As an evaluation metric to measure the performance of a prediction model, a forecast error is computed from the difference between observed values and predicted values, that is, the ETF closing prices of the testing dataset vs. the predicted ETF closing prices. Prevalent forecast errors, such as MAE and RMSE, are scale dependent and cannot compare the time series of different units. As a way around scale dependency, scaled errors \cite{Hyndman2006} compute the forecast errors using the MAE attained during training of the forecasting model. In this study, forecasting is performed on nonseasonal time series and scaled errors are computed using naïve forecasts, as provided in \cite{Hyndman2021}. Scaled errors used to evaluate machine learning model performance during training are the mean absolute scaled error (MASE) and the root mean squared scaled error (RMSSE). A scaled error is (favourably) lower than one if forecasts are more accurate than the average one-step naïve forecast computed on the training set. Scaled errors are computed as follows:
\begin{center}
\begin{equation}\label{MASE}
MASE=mean(|q_{j}|)
\end{equation}
\begin{equation}\label{RMSSE}
RMSSE=\sqrt{mean(q_{j}^{2})}
\end{equation}
\end{center}
where $q_{j}$ is:
\begin{center}
\begin{equation}\label{qj}
q_{j}=\frac{e_{j}}{\frac{1}{T-1}\sum_{t=2}^{T} |y_{t}-y_{t-1}|}
\end{equation}
\end{center}
and $q_{j}^{2}$ is the square of the aforementioned $q_{j}$, $e_{j}$ is the forecast error, $\{y_1,...,y_t\}$ is the training data, and $T$ is the size of the training set.

To determine if the findings from the portfolio optimisation module are consistent, the optimisation algorithm is run multiple times using different (randomised) combinations of 100 ETFs.  In every run, the expected annualized portfolio return, annualized risk, Sharpe ratio \cite{Sharpe1966}, and average portfolio ESG-rating are computed. Sharpe ratios are computed to relate the return of an investment to the risk of such an investment, where a favourable Sharpe ratio is greater than 1.

The optimisation function for the traditional MV model is calculated as a minimising function of the negative Sharpe ratio ($S_p$);
\begin{center}
\begin{equation}\label{maxsharpe}
min(-S_{p})=min(-\frac{r_{p}-r_{f}}{\sigma_{p}})
\end{equation}
\end{center}
where $r_{f}$ is the risk-free return and portfolio return ($r_{p}$) is
\begin{center}
\begin{equation}\label{return}
r_{p}=\sum_{i=1}^{N}w_{i}r_{i}
\end{equation}
\end{center}
and portfolio risk ($\sigma_{p}$) is,
\begin{center}
\begin{equation}\label{risk}
\sigma_{p}=\sum_{i=1}^{N}\sum_{j=1}^{N}w_{i}\sigma_{ij}w_{j}
\end{equation}
\end{center}
and $w_{i}$ is the weight allocated to the ETF in the portfolio. Similarly, the ESG-MV model is calculated as a minimising function of the negative Sharpe ratio ($\tilde{S}_{p}$) but now includes the average portfolio ESG-rating, which is the mean of weighted ESG scores of individual ETFs; 
\begin{center}
\begin{equation}\label{esgsharpe}
min(-\tilde{S}_{p})=min(-ESG_{p}\frac{r_{p}-r_{f}}{\sigma_{p}})
\end{equation}
\end{center}

Due to the available data on the ESG ratings, we cannot use historical ESG ratings to improve the precision of our portfolio optimisation, as was done in \citet{Vo2019}.

\subsection{Tools and Libraries}
All experiments were run on a 1.8GHz Dual-Core Intel Core i5 (5350U). All algorithms were implemented using Python 3.3 via the Jupyter Notebook open-source IDE. Python libraries used to implement the algorithms in this study include NumPy, SciPy, Statistics, Math, Matplotlib, Seaborn, Pandas, Tensorflow, and Scikit-Learn.

All related code, notebooks, datasets, and results can be viewed \href{https://github.com/Taeish/Machine-Learning-for-Socially-Responsible-Portfolio-Optimisation}{here}.

\section{Results \& Discussion} \label{section:results}
\subsection{Results of the prediction model}
\begin{figure}[htb]
    \centering
    \includegraphics[width=0.8\columnwidth]{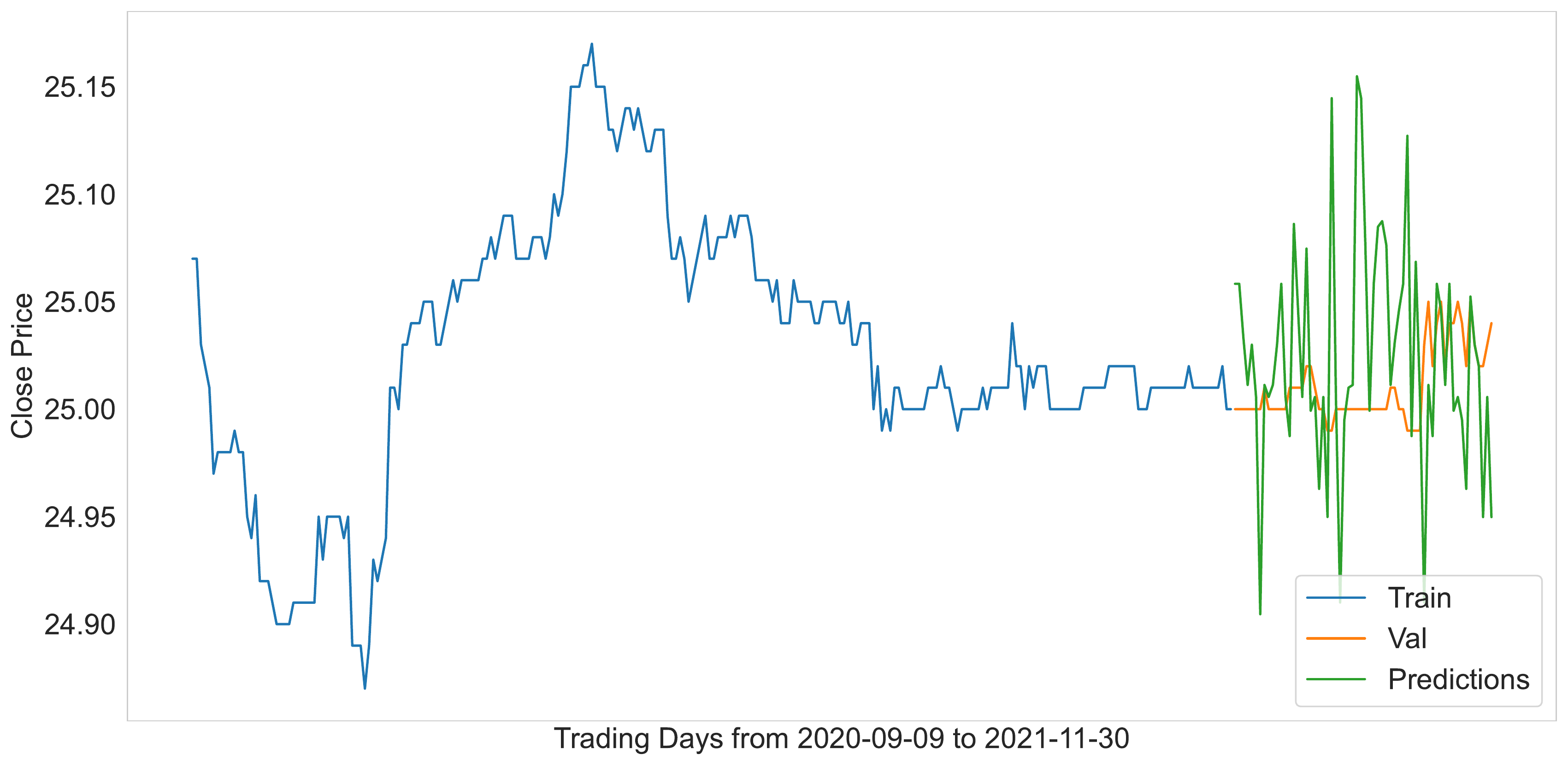}
    \caption{Closing Prices using Random Forest Model}
    \label{fig:RF val}
\end{figure}

\begin{figure}[htb]
    \centering
    \includegraphics[width=0.8\columnwidth]{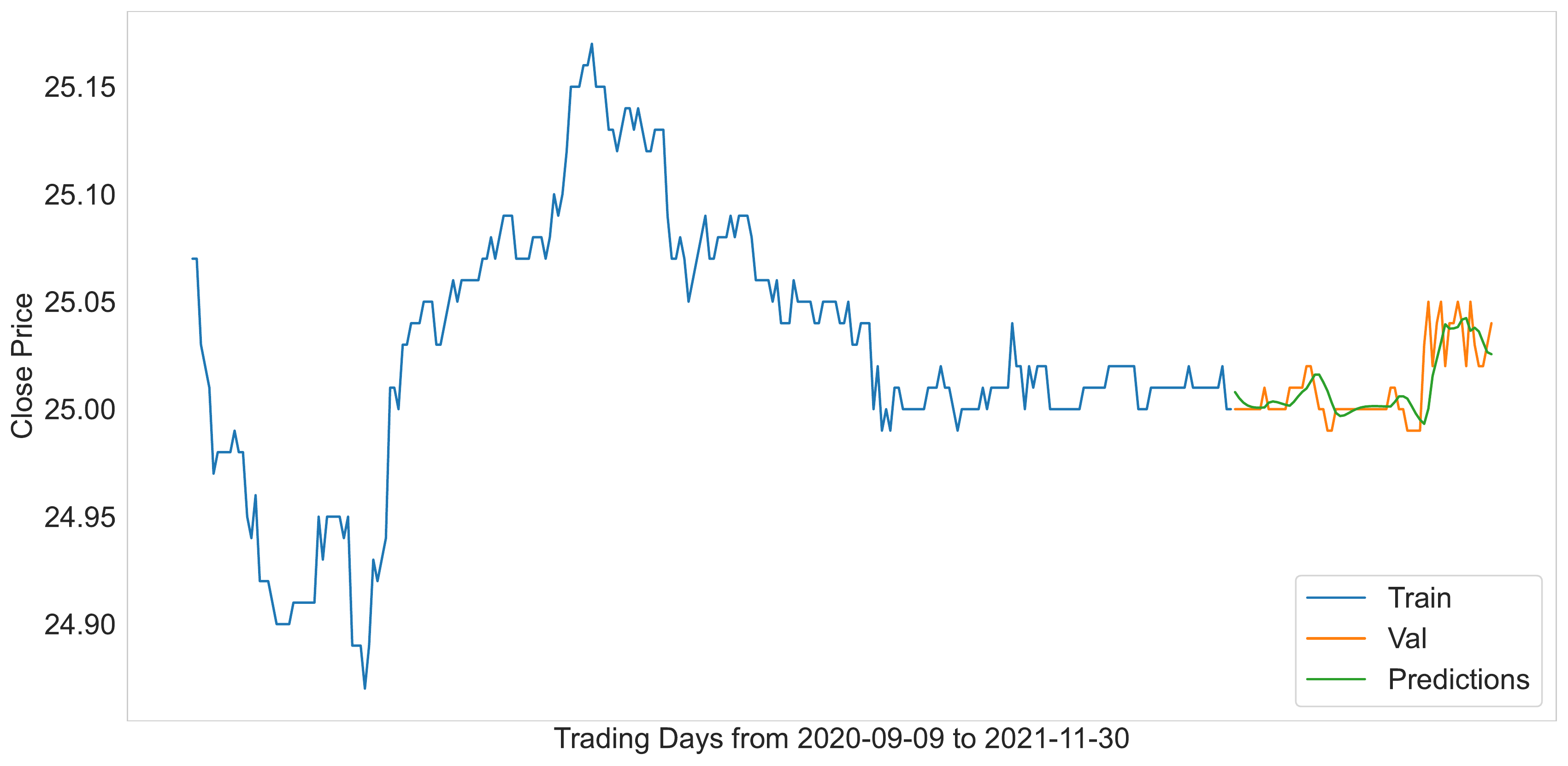}
    \caption{Closing Prices using LSTM Model}
    \label{fig:LSTM val}
\end{figure}

\begin{figure}[htb]
    \centering
    \includegraphics[width=0.8\columnwidth]{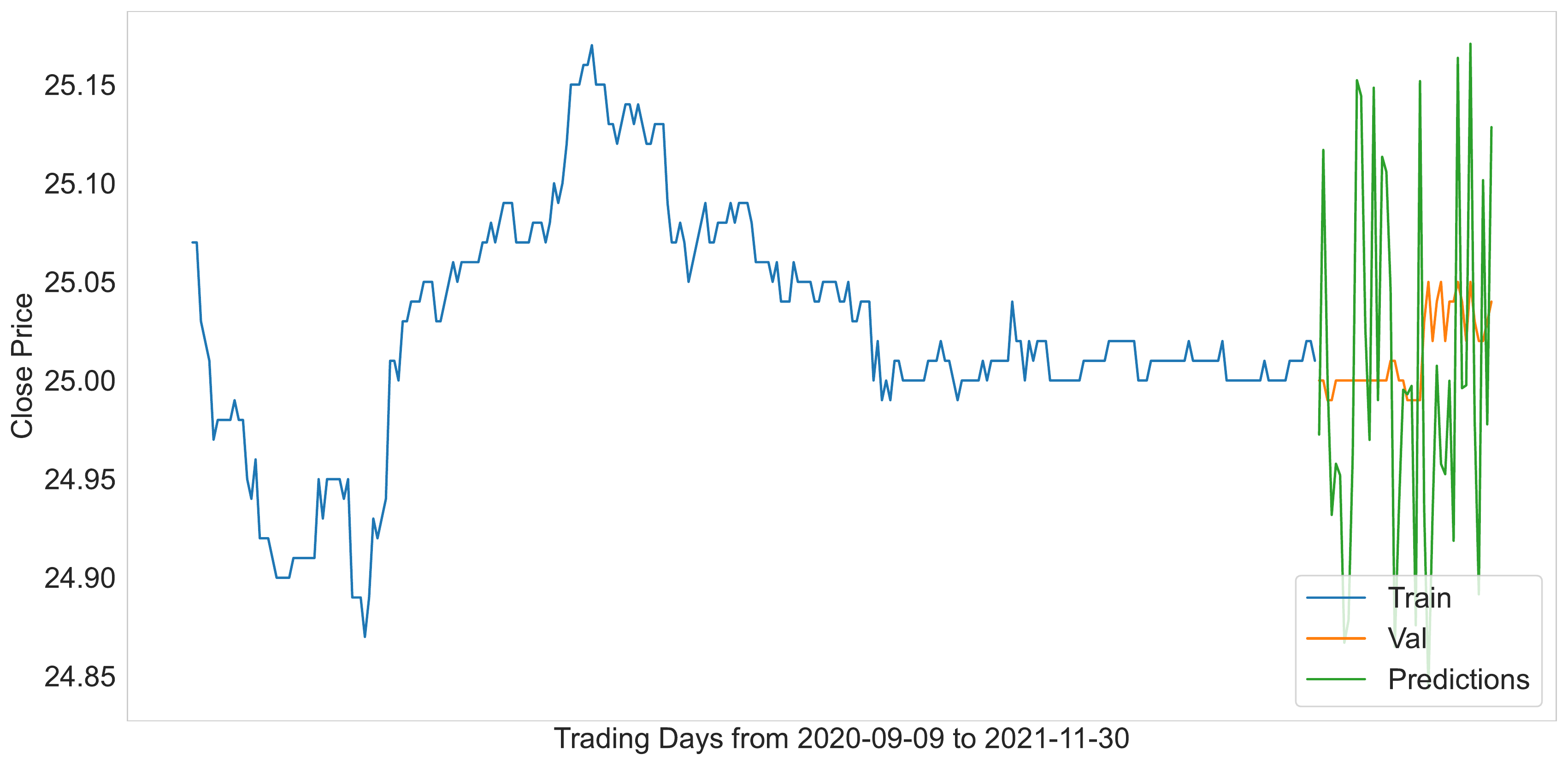}
    \caption{Closing Prices using CNN Model}
    \label{fig:CNN val}
\end{figure}

Adjusted closing prices for AAA ETF are predicted using machine learning algorithms. Figures \ref{fig:RF val} to \ref{fig:CNN val} show the performance during training and testing. 


\begin{figure}[htb]
    \centering
    \includegraphics[width=0.8\columnwidth]{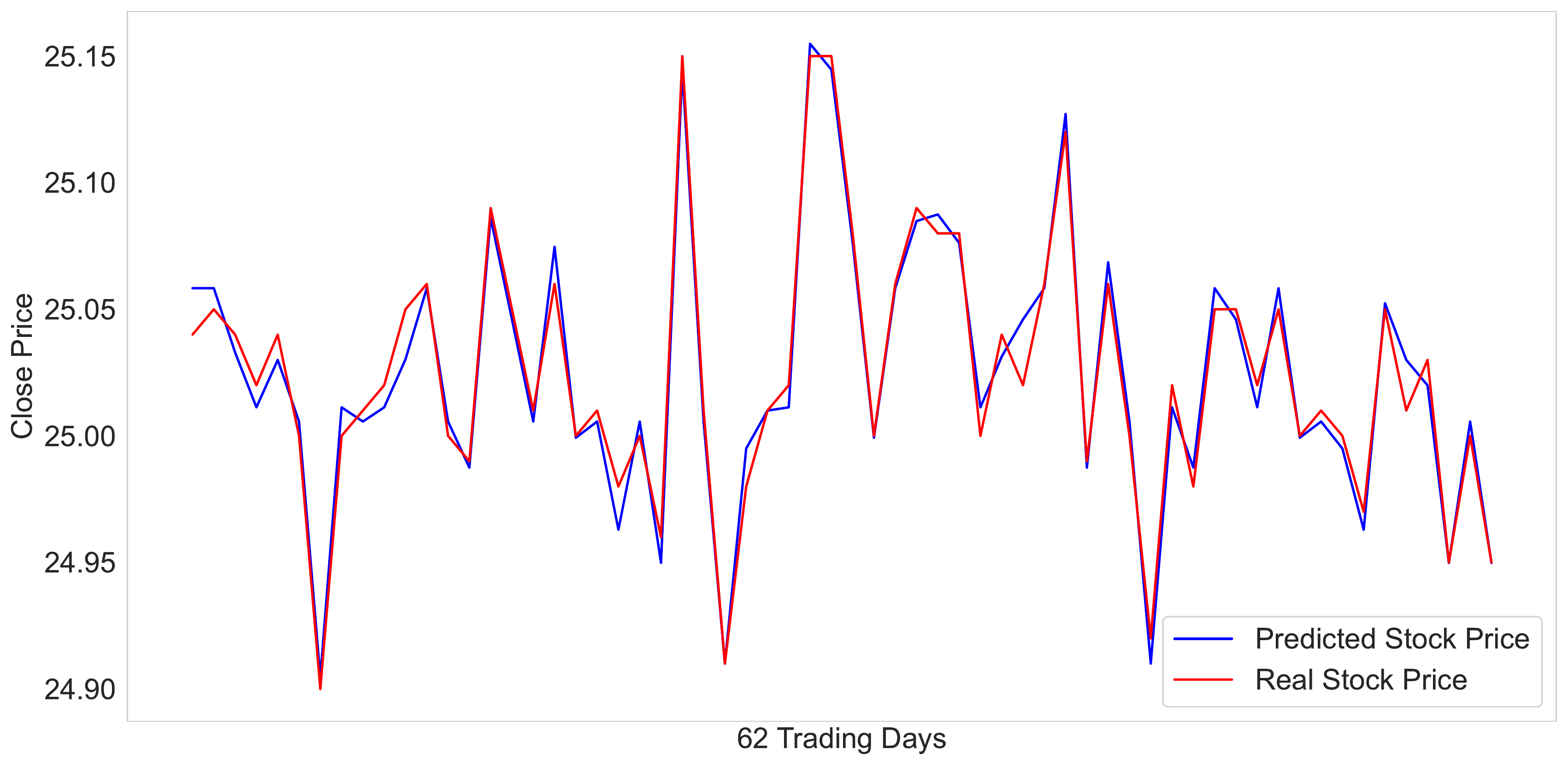}
    \caption{Training and Validation using Random Forest Model}
    \label{fig:RF train}
\end{figure}

\begin{figure}[htb]
    \centering
    \includegraphics[width=0.8\columnwidth]{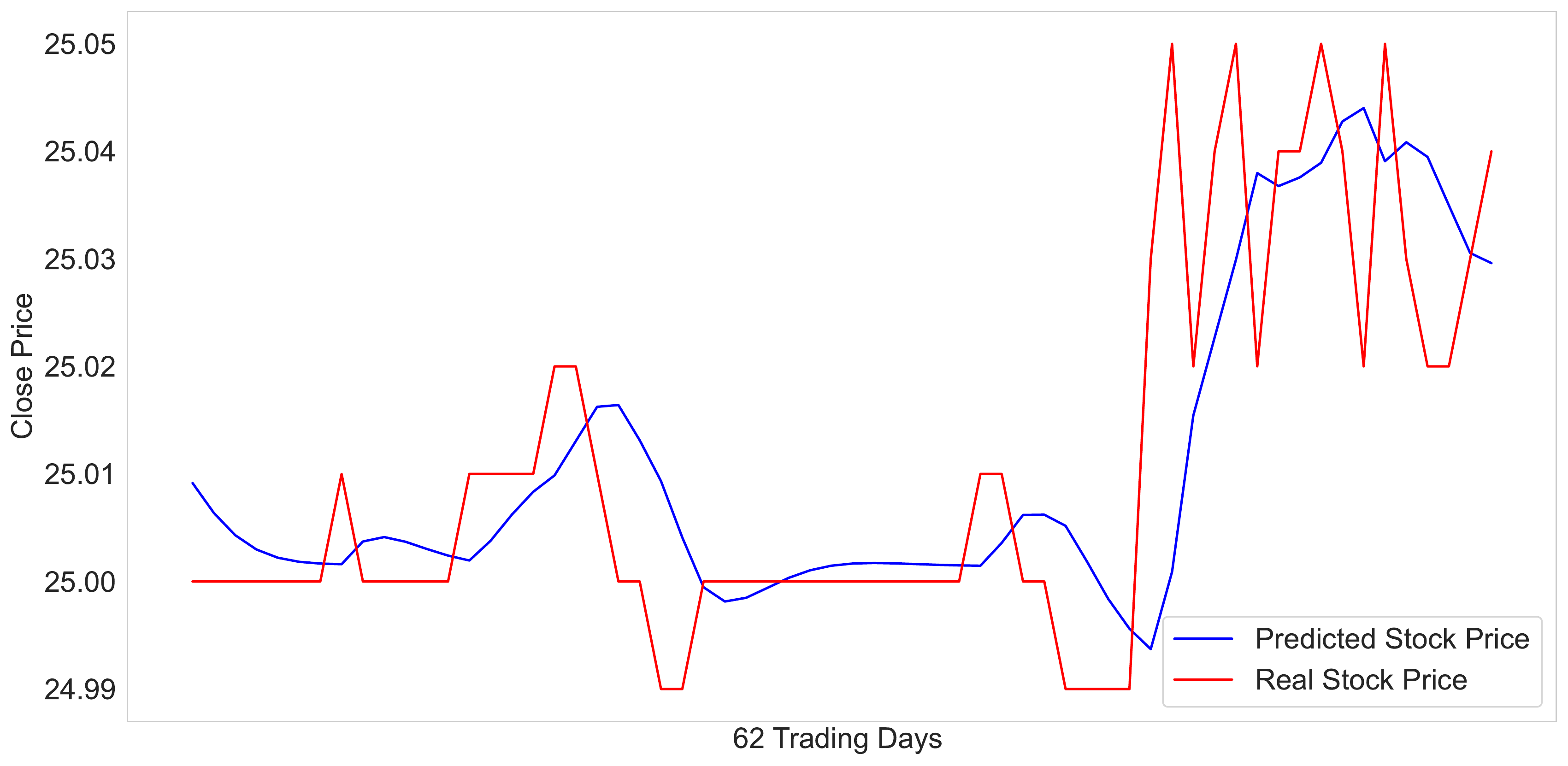}
    \caption{Training and Validation using LSTM Model}
    \label{fig:LSTM train}
\end{figure}

\begin{figure}[htb]
    \centering
    \includegraphics[width=0.8\columnwidth]{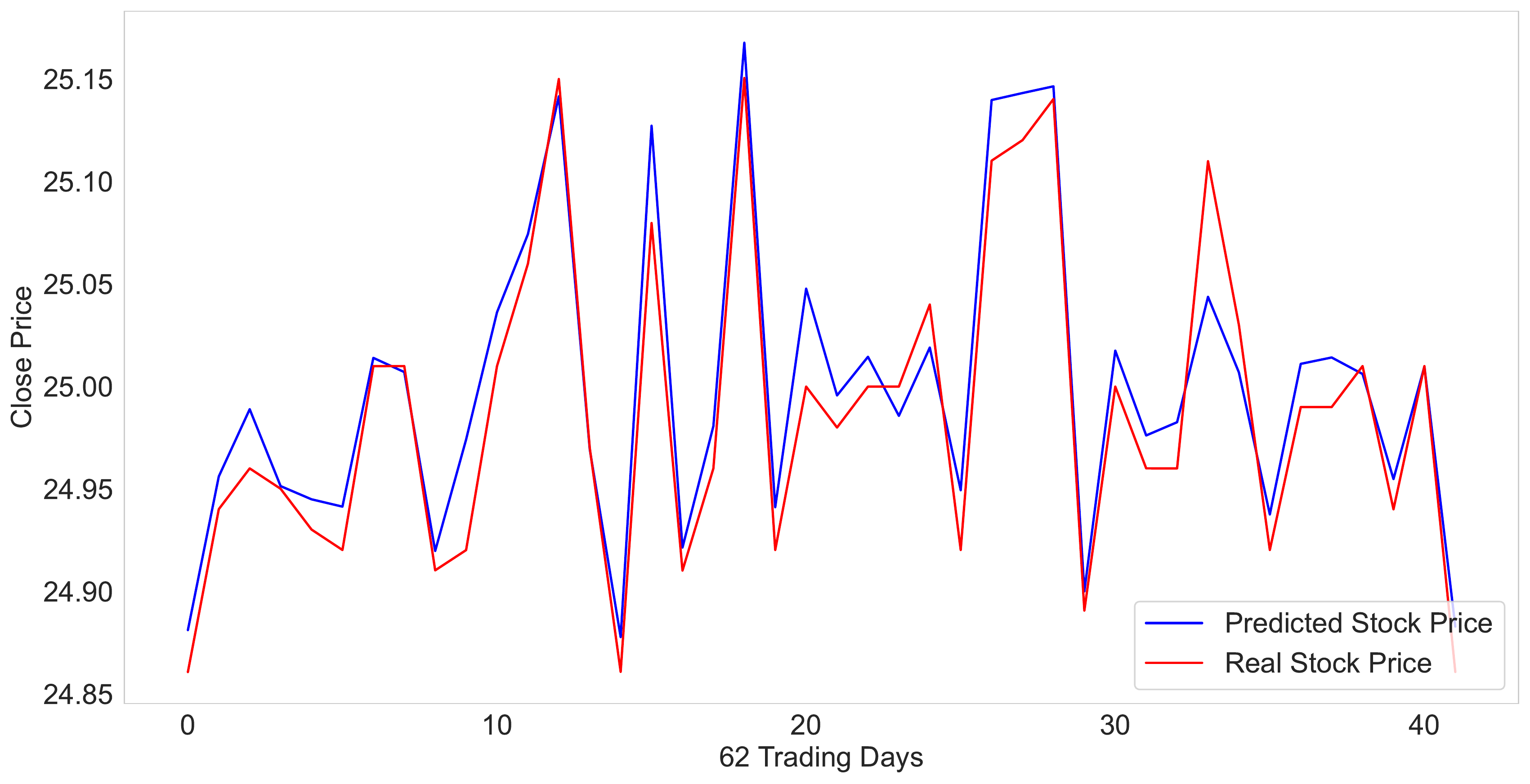}
    \caption{Training and Validation using CNN Model}
    \label{fig:CNN train}
\end{figure}

It can be seen from Figures \ref{fig:RF val}, \ref{fig:LSTM val}, and \ref{fig:CNN val}, during testing and validation, the RF and CNN models show high variance (i.e. overfitting) whilst the LSTM shows underfitting. To specifically investigate the behaviour of the prediction models during training, Figures \ref{fig:RF train} to \ref{fig:CNN train} plot the actual and predicted data, omitting the test data. When closely looking at Figure \ref{fig:RF train}, the RF model overestimates the troughs and peaks. Figure \ref{fig:LSTM train} shows that the LSTM model is significantly under-fitting. Figure \ref{fig:CNN train} shows that CNN does the opposite of RF and underestimates troughs and peaks. 


Tables \ref{table:AAA ETF} to \ref{table:UOCT ETF} provide the MASE and MAE of the models when predicting the adjusted closing prices of the AAA, SPY and UOCT ETF.

\begin{table}[htb]
    \caption{MASE and RMSE when predicting AAA ETF closing prices}
    \centering
        \begin{tabular}{lrrr}
        \toprule
             Model&MASE&RMSSE&Confidence Interval\\
             \bottomrule
               
            RF      &   0.0014    &	0.0011 & (25.0110, 25.0372) 
            \\LSTM    &   0.0126	 &0.0077&(25.0123, 25.0184)
            \\CNN     &   0.3105	&0.2500&(24.9973, 25.0468) \\
            \bottomrule
        \end{tabular}
    \label{table:AAA ETF}
\end{table}

\begin{table}[htb]
\caption{MASE and RMSE when predicting SPY ETF closing prices}
    \centering
        \begin{tabular}{lrrr}
        \toprule
             Model&MASE&RMSSE&Confidence Interval\\
             \bottomrule
            RF      &   0.0000 & 0.0000& (148.93, 157.78)
            \\LSTM    &   706.5726	 & 479.5685 &(283.95, 289.69)
            \\CNN     &   0.0445	&0.0337&(284.26, 291.01) \\
             \bottomrule
        \end{tabular}
    \label{table:SPY ETF}
\end{table}
\begin{table}[htb]
    \caption{MASE and RMSE when predicting UOCT ETF closing prices}
    \centering
        \begin{tabular}{lrrr}
        \toprule
             Model&MASE&RMSSE&Confidence Interval\\
             \bottomrule           
            RF      &   0.001315 & 0.001055 & (25.3791, 25.8541)
            \\LSTM    &  0.337371 & 0.209528 & (27.7632, 27.8420)
            \\CNN     &   0.040433 & 0.030315 & (27.6718, 27.9850) \\
             \bottomrule
        \end{tabular}
    \label{table:UOCT ETF}
\end{table}

From Table \ref{table:SPY ETF}, the LSTM model performs poorly with wider confidence intervals. This could be attributed to the underfitting due to the model architecture. 

To improve the performance of the three models, effective regularisation techniques need to be incorporated into the models to reduce generalisation errors. If implemented correctly, regularisation will improve the variance issues displayed in the RF and CNN models. The CNN model could be too complex for this time series model, leading to high variance.

\begin{figure}[htb]
    \centering
    \includegraphics[width=0.8\columnwidth]{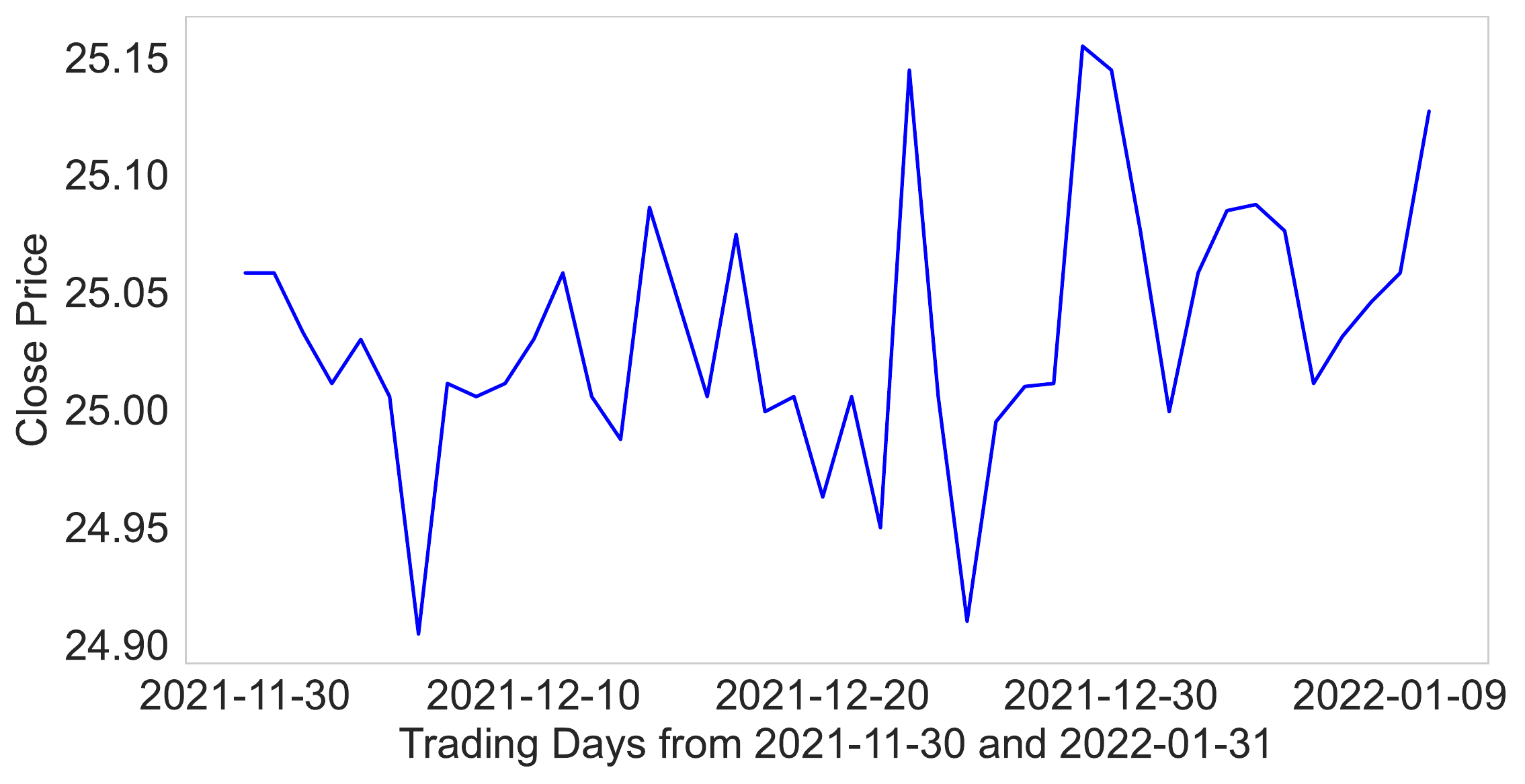}
    \caption{Forecast Adjusted Closing Prices of AAA ETF}
    \label{fig:RFpred}
\end{figure}

Regardless of expected ETF prices, the RF outperforms the other models by having the lowest MASE and RMSE. Due to this, the RF model is used to predict the adjusted closing prices of ETFs for the 2021-11-30 to 2022-01-31 trading period. Figure \ref{fig:RFpred} plots the forecasting of the AAA ETF for the 2021-11-30 to 2022-01-31 trading period using the RF model. This is a sample of the forecasts produced, stored and utilised in the portfolio optimisation component of the study.

\subsection {Portfolio optimisation model results}

\begin{table}[htb]
    \caption{Tabulated Results from Optimisation Module}
    \centering
        \begin{tabular}{p{0.3in}p{0.5in}p{1in}p{1in}p{1in}p{1in}}
        \toprule
Run	&	
Model	&	
Sharpe Ratio &	
Risk	&	
Return	&	
Mean ESG Score \\
\bottomrule
	1	&	MV 	&	2.174	&	11.486	&	31.218	&	5.204
\\		&	ESG MV	&	1.981	&	10.062	&	26.182	&	7.131
\\	2	&	MV 	&	2.031	&	6.265	&	18.974	&	5.725
\\		&	ESG MV	&	2.863	&	5.440	&	21.823	&	7.208 
\\	3	&	MV 	&	2.475	&	9.441	&	29.617	&	5.899
\\		&	ESG MV	&	2.213	&	8.687	&	25.470	&	7.878
\\	4	&	MV 	&	1.933	&	8.751	&	23.166	&	1.715
\\		&	ESG MV	&	1.283	&	5.836	&	13.736	&	5.787
\\	5	&	MV 	&	1.831	&	9.447	&	23.545	&	5.903
\\		&	ESG MV	&	1.623	&	6.613	&	16.981	&	7.537
\\	6	&	MV 	&	3.277	&	9.430	&	37.156	&	6.375
\\		&	ESG MV	&	3.039	&	7.503	&	29.052	&	7.294
\\	7	&	MV 	&	1.679	&	12.03	&	26.445	&	3.584
\\		&	ESG MV	&	1.330	&	7.332	&	16.004	&	6.069
\\	8	&	MV 	&	1.589	&	12.58	&	26.242	&	3.044
\\		&	ESG MV	&	1.097	&	6.234	&	13.086	&	6.375
\\	9	&	MV 	&	2.120	&	5.550	&	18.017	&	6.179
\\		&	ESG MV	&	1.990	&	6.140	&	18.465	&	7.307
\\	10	&	MV 	&	3.172	&	6.482	&	26.813	&	5.555
\\		&	ESG MV	&	2.955	&	6.420	&	25.218	&	7.252
\\	11	&	MV 	&	1.920	&	4.841	&	15.546	&	4.300
\\		&	ESG MV	&	1.541	&	4.775	&	13.607	&	6.483
\\	12	&	MV 	&	2.294	&	7.709	&	23.933	&	2.662
\\		&	ESG MV	&	1.554	&	4.475	&	13.203	&	6.238\\
             \hline
        \end{tabular}    
    \label{table:PO}
\end{table}

The results obtained from portfolio optimisation using the MV and ESG-MV methods are provided in Table \ref{table:PO}. The table shows results from a market size of 100, where the market is randomly generated in every run (that is, every run contains a random selection of 100 ETFs to optimise over). Note that all values in the table are annualised. 

When comparing the results of the MV model to the ESG-MV models, the following are observed:
\begin{itemize}
    \item There is an average decrease of 12.29\% in the Sharpe ratio.  SR portfolios perform slightly worse on average at the expense of achieving a drastic increase of 65.72\% in mean portfolio ESG scores.
    \item There is an average decrease of 20\% in annualised risk and an average reduction of 21.32\% in annualised return. This implies that SR portfolios are less risky investments, but investors (as expected) suffer a lower return from said portfolios.
\end{itemize}
The ESG-MV model is still capable of achieving positive returns. This is achieved by maximizing the Sharpe Ratio in traditional MV models, which optimally selects ETFs with higher returns and avoids negative annualised returns. All Sharpe Ratios are favourably greater than 1, indicating that all optimal portfolio weights produced in the experiment have resulted in attractive investment portfolios.

In run two from Table \ref{table:PO}, the Sharpe ratio using ESG-MV is higher than that of the MV model. This indicates that the ESG-MV model can achieve higher ESG ratings and Sharpe ratios than the MV model in particular markets. This shows that SR portfolios can have competitive financial performance compared to portfolios produced from the traditional MV model.

\begin{table}[htb]
\caption{Asset allocation weights from MV optimisation vs MV-ESG optimisation, sorted by ascending ESG scores}
    \centering
        \begin{tabular}{p{1in}p{1in}p{1in}p{1in}
        }
      
        \toprule
ETF	&	
ESG Score	&MV Optimal weights (\%)	&	
ESG-MV Optimal weights (\%)	\\ 
\bottomrule
	 SMMU	&	0	&	0	&	0	
\\	 SDP	&	0	&	6.624	&	0	
\\	 PAUG	&	0	&	0	&	0	
\\
     IBMQ	&	0	&	0	&	0	
\\	 FUMB	&	0	&	0	&	0	
\\	 HIPS	&	0	&	2.42	&	0	
\\	 DJUL	&	0	&	0	&	0	
\\	 XBJL	&	0	&	0	&	0	
\\	 BNKU	&	0	&	3.433	&	0	
\\
     MLPA	&	3.73	&	7.052	&	2.324	
\\	 OSCV	&	3.97	&	0	&	0	
\\
     PSQ	&	5.81	&	0.158	&	0.012	
\\	 SCHQ	&	5.81	&	0	&	0	
\\	 SDOW	&	5.81	&	27.412	&	25.073	
\\	 SPTS	&	5.81	&	0	&	0	
\\
     FNDB	&	7.66	&	7.756	&	4.894	
\\	 IVE	&	7.68	&	0.485	&	5.225	
\\	 VOOV	&	7.68	&	0.057	&	0.003	
\\	 FNCL	&	7.72	&	7.779	&	4.875	
\\	 LRNZ	&	7.79	&	0	&	0	
\\	 SPUU	&	7.88	&	4.102	&	3.259	
\\	 EPS	&	7.89	&	0	&	0	
\\
     USXF	&	9.22	&	0	&	2.305	
\\	 WBIG	&	9.38	&	0	&	13.372	
\\	 RBND	&	9.57	&	0	&	0	\\
\bottomrule
        \end{tabular}
    \label{table:weights}
\end{table}

Table \ref{table:weights} provides the weights allocated to each of the 100 ETFs in the market generated during run 5 of the experiment. From Table \ref{table:weights}, it is seen that ESG scores drive the optimal weights of ETFs with high ESG ratings (see WBIG ETF in Table \ref{table:weights}).

In this particular run, both models consist of 33 non-zero weighted ETFs. However, zero-weighted ETFs in the MV model do not correlate with zero-weighted ETFs in ESG-MV. This is attributed to a higher ESG score being traded off by the risk and return of the ETF. This can be seen in the change in weightings of WBIG and SDP ETF. More specifically, WBIG (with a high ESG score of 9.38) is assigned 13.37\% in the ESG-MV model but 0\% in the MV model, while SDP (with a low ESG score of 0) is assigned 0\% in the ESG-MV model but 6.62\% in the MV model.
\section{Conclusion} \label{section:conc}
This study was designed to leverage machine learning in the computation of optimal SRI portfolios to evaluate the effect of considering ESG ratings in portfolio optimisation. Random forest, LSTM neural network, and CNN models were implemented to predict and forecast adjusted closing prices of selected ETFs. In evaluating model performance, the RF model outperformed the deep learning models during model testing. Consequently, the RF model was used to forecast the adjusted closing values of all ETFs.

Using historical and forecasted data, this study has shown that ESG-MV models produce competitive portfolios with a slight trade-off between a lower Sharpe ratio for higher mean ESG scores of the portfolio. The findings will interest SR investors seeking a portfolio with a balance of moral and economic value. The study is limited by the absence of historical data on ESG scores as utilised in \citet{Vo2019}. The study should be repeated using regularisation in machine learning models to improve the overall performance of the models' forecasting accuracy.

\begin{acks}
We thank the Nedbank Research Chair for his guidance, dedication, and invaluable insights while writing this paper. To my friends and family for their unwavering support and encouragement.
\end{acks}

\bibliographystyle{ACM-Reference-Format}
\bibliography{sample-base}


\begin{thebibliography}{16}


\ifx \showCODEN    \undefined \def \showCODEN     #1{\unskip}     \fi
\ifx \showDOI      \undefined \def \showDOI       #1{#1}\fi
\ifx \showISBNx    \undefined \def \showISBNx     #1{\unskip}     \fi
\ifx \showISBNxiii \undefined \def \showISBNxiii  #1{\unskip}     \fi
\ifx \showISSN     \undefined \def \showISSN      #1{\unskip}     \fi
\ifx \showLCCN     \undefined \def \showLCCN      #1{\unskip}     \fi
\ifx \shownote     \undefined \def \shownote      #1{#1}          \fi
\ifx \showarticletitle \undefined \def \showarticletitle #1{#1}   \fi
\ifx \showURL      \undefined \def \showURL       {\relax}        \fi
\providecommand\bibfield[2]{#2}
\providecommand\bibinfo[2]{#2}
\providecommand\natexlab[1]{#1}
\providecommand\showeprint[2][]{arXiv:#2}

\bibitem[Branch et~al\mbox{.}(2019)]%
        {branch2019guide}
\bibfield{author}{\bibinfo{person}{Michael Branch}, \bibinfo{person}{Lisa~R
  Goldberg}, {and} \bibinfo{person}{Pete Hand}.}
  \bibinfo{year}{2019}\natexlab{}.
\newblock \showarticletitle{A guide to ESG portfolio construction}.
\newblock \bibinfo{journal}{\emph{The Journal of Portfolio Management}}
  \bibinfo{volume}{45}, \bibinfo{number}{4} (\bibinfo{year}{2019}),
  \bibinfo{pages}{61--66}.
\newblock


\bibitem[Chen et~al\mbox{.}(2021)]%
        {chen2021social}
\bibfield{author}{\bibinfo{person}{Li Chen}, \bibinfo{person}{Lipei Zhang},
  \bibinfo{person}{Jun Huang}, \bibinfo{person}{Helu Xiao}, {and}
  \bibinfo{person}{Zhongbao Zhou}.} \bibinfo{year}{2021}\natexlab{}.
\newblock \showarticletitle{Social responsibility portfolio optimization
  incorporating ESG criteria}.
\newblock \bibinfo{journal}{\emph{Journal of Management Science and
  Engineering}} \bibinfo{volume}{6}, \bibinfo{number}{1}
  (\bibinfo{year}{2021}), \bibinfo{pages}{75--85}.
\newblock


\bibitem[De~Spiegeleer et~al\mbox{.}(2021)]%
        {de2021esg}
\bibfield{author}{\bibinfo{person}{Jan De~Spiegeleer}, \bibinfo{person}{Stephan
  H{\"o}cht}, \bibinfo{person}{Daniel Jakubowski}, \bibinfo{person}{Sofie
  Reyners}, {and} \bibinfo{person}{Wim Schoutens}.}
  \bibinfo{year}{2021}\natexlab{}.
\newblock \showarticletitle{ESG: A new dimension in portfolio allocation}.
\newblock \bibinfo{journal}{\emph{Journal of Sustainable Finance \&
  Investment}}  \bibinfo{volume}{1} (\bibinfo{year}{2021}),
  \bibinfo{pages}{1--41}.
\newblock


\bibitem[Freeborough and van Zyl(2022)]%
        {freeborough2022investigating}
\bibfield{author}{\bibinfo{person}{Warren Freeborough} {and}
  \bibinfo{person}{Terence~L van Zyl}.} \bibinfo{year}{2022}\natexlab{}.
\newblock \showarticletitle{Investigating Explainability Methods in Recurrent
  Neural Network Architectures for Financial Time Series Data}.
\newblock \bibinfo{journal}{\emph{Applied Sciences}} \bibinfo{volume}{12},
  \bibinfo{number}{3} (\bibinfo{year}{2022}), \bibinfo{pages}{1427}.
\newblock


\bibitem[Hyndman and Athanasopoulos(2021)]%
        {Hyndman2021}
\bibfield{author}{\bibinfo{person}{R.~J. Hyndman} {and} \bibinfo{person}{G.
  Athanasopoulos}.} \bibinfo{year}{2021}\natexlab{}.
\newblock \bibinfo{booktitle}{\emph{Forecasting: principles and practice}
  (\bibinfo{edition}{3} ed.)}.
\newblock \bibinfo{publisher}{OTexts}, \bibinfo{address}{Australia}.
\newblock


\bibitem[Hyndman and Koehler(2006)]%
        {Hyndman2006}
\bibfield{author}{\bibinfo{person}{R.~J. Hyndman} {and} \bibinfo{person}{A.B.
  Koehler}.} \bibinfo{year}{2006}\natexlab{}.
\newblock \showarticletitle{Another look at measures of forecast accuracy}.
\newblock \bibinfo{journal}{\emph{International Journal of Forecasting}}
  \bibinfo{volume}{22}, \bibinfo{number}{4} (\bibinfo{year}{2006}),
  \bibinfo{pages}{679–688}.
\newblock


\bibitem[Laher et~al\mbox{.}(2021)]%
        {laher2021deep}
\bibfield{author}{\bibinfo{person}{Siddeeq Laher}, \bibinfo{person}{Andrew
  Paskaramoorthy}, {and} \bibinfo{person}{Terence~L Van~Zyl}.}
  \bibinfo{year}{2021}\natexlab{}.
\newblock \showarticletitle{Deep learning for financial time series forecast
  fusion and optimal portfolio rebalancing}. In \bibinfo{booktitle}{\emph{2021
  IEEE 24th International Conference on Information Fusion (FUSION)}}. IEEE,
  \bibinfo{publisher}{IEEE}, \bibinfo{address}{South Africa},
  \bibinfo{pages}{1--8}.
\newblock


\bibitem[Markowitz(1952)]%
        {Markowitz1952}
\bibfield{author}{\bibinfo{person}{Harry Markowitz}.}
  \bibinfo{year}{1952}\natexlab{}.
\newblock \showarticletitle{Portfolio Selection}.
\newblock \bibinfo{journal}{\emph{Journal of Finance}} \bibinfo{volume}{7},
  \bibinfo{number}{1} (\bibinfo{year}{1952}), \bibinfo{pages}{77--91}.
\newblock


\bibitem[Muthivhi and van Zyl(2022)]%
        {muthivhi2022fusion}
\bibfield{author}{\bibinfo{person}{Mufhumudzi Muthivhi} {and}
  \bibinfo{person}{Terence~L van Zyl}.} \bibinfo{year}{2022}\natexlab{}.
\newblock \showarticletitle{Fusion of Sentiment and Asset Price Predictions for
  Portfolio Optimization}. In \bibinfo{booktitle}{\emph{2022 25th International
  Conference on Information Fusion (FUSION)}}. IEEE, \bibinfo{publisher}{IEEE},
  \bibinfo{address}{Sweden}, \bibinfo{pages}{1--8}.
\newblock


\bibitem[Ouchen(2022)]%
        {ouchen2022esg}
\bibfield{author}{\bibinfo{person}{Abdessamad Ouchen}.}
  \bibinfo{year}{2022}\natexlab{}.
\newblock \showarticletitle{Is the ESG portfolio less turbulent than a market
  benchmark portfolio?}
\newblock \bibinfo{journal}{\emph{Risk Management}} \bibinfo{volume}{24},
  \bibinfo{number}{1} (\bibinfo{year}{2022}), \bibinfo{pages}{1--33}.
\newblock


\bibitem[Pai(2018)]%
        {Pai2018}
\bibfield{author}{\bibinfo{person}{G.~A.~Vijayalakshmi Pai}.}
  \bibinfo{year}{2018}\natexlab{}.
\newblock \bibinfo{booktitle}{\emph{Metaheuristics for Portfolio Optimization-
  An Introduction using MATLAB}}.
\newblock \bibinfo{publisher}{Wiley-ISTE}, \bibinfo{address}{London}.
\newblock


\bibitem[Paskaramoorthy et~al\mbox{.}(2020)]%
        {paskaramoorthy2020framework}
\bibfield{author}{\bibinfo{person}{Andrew~B Paskaramoorthy},
  \bibinfo{person}{Tim~J Gebbie}, {and} \bibinfo{person}{Terence~L van Zyl}.}
  \bibinfo{year}{2020}\natexlab{}.
\newblock \showarticletitle{A framework for online investment decisions}.
\newblock \bibinfo{journal}{\emph{Investment Analysts Journal}}
  \bibinfo{volume}{49}, \bibinfo{number}{3} (\bibinfo{year}{2020}),
  \bibinfo{pages}{215--231}.
\newblock


\bibitem[Sharpe(1966)]%
        {Sharpe1966}
\bibfield{author}{\bibinfo{person}{William~F Sharpe}.}
  \bibinfo{year}{1966}\natexlab{}.
\newblock \showarticletitle{Mutual fund performance}.
\newblock \bibinfo{journal}{\emph{The Journal of Business}}
  \bibinfo{volume}{39}, \bibinfo{number}{1} (\bibinfo{year}{1966}),
  \bibinfo{pages}{119--138}.
\newblock


\bibitem[Skeepers et~al\mbox{.}(2021)]%
        {skeepers2021ma}
\bibfield{author}{\bibinfo{person}{Tarrin Skeepers}, \bibinfo{person}{Terence~L
  van Zyl}, {and} \bibinfo{person}{Andrew Paskaramoorthy}.}
  \bibinfo{year}{2021}\natexlab{}.
\newblock \showarticletitle{MA-FDRNN: Multi-asset fuzzy deep recurrent neural
  network reinforcement learning for portfolio management}. In
  \bibinfo{booktitle}{\emph{2021 8th International Conference on Soft Computing
  \& Machine Intelligence (ISCMI)}}. IEEE, \bibinfo{publisher}{IEEE},
  \bibinfo{address}{Cairo, Egypt}, \bibinfo{pages}{32--37}.
\newblock


\bibitem[van Zyl et~al\mbox{.}(2021)]%
        {van2021parden}
\bibfield{author}{\bibinfo{person}{Terence~L van Zyl}, \bibinfo{person}{Matthew
  Woolway}, {and} \bibinfo{person}{Andrew Paskaramoorthy}.}
  \bibinfo{year}{2021}\natexlab{}.
\newblock \showarticletitle{Parden: Surrogate assisted hyper-parameter
  optimisation for portfolio selection}. In \bibinfo{booktitle}{\emph{2021 8th
  international conference on soft computing \& machine intelligence (ISCMI)}}.
  IEEE, \bibinfo{publisher}{IEEE}, \bibinfo{address}{Cairo, Egypt},
  \bibinfo{pages}{101--107}.
\newblock


\bibitem[Vo et~al\mbox{.}(2019)]%
        {Vo2019}
\bibfield{author}{\bibinfo{person}{Nhi~NY Vo}, \bibinfo{person}{Xuezhong He},
  \bibinfo{person}{Shaowu Liu}, {and} \bibinfo{person}{Guandong Xu}.}
  \bibinfo{year}{2019}\natexlab{}.
\newblock \showarticletitle{Deep learning for decision making and the
  optimization of socially responsible investments and portfolio}.
\newblock \bibinfo{journal}{\emph{Decision Support Systems}}
  \bibinfo{volume}{124} (\bibinfo{year}{2019}), \bibinfo{pages}{113097}.
\newblock


\end{thebibliography}

\end{document}